\documentclass[conference]{IEEEtran}
\IEEEoverridecommandlockouts
\usepackage{cite}
\usepackage{amsmath,amssymb,amsfonts}
\usepackage{algorithmic}
\usepackage{textcomp}
\usepackage{booktabs}
\usepackage{lipsum}
\usepackage{hyperref}  	% Creates hyperlinks in cross references
\hypersetup{
    allcolors=black,
    colorlinks=true,
    linkcolor=black,
    urlcolor=blue,
}

\usepackage[dvipsnames]{xcolor}
\usepackage{graphicx}
\usepackage{tikz}
\usepackage[printwatermark]{xwatermark}
\newsavebox\mybox
\savebox\mybox{\tikz[color=black,opacity=0.3]\node{PREPRINT};}
\newwatermark*[
  allpages,
  angle=55,
  scale=17,
  xpos=-70,
  ypos=50
]{\usebox\mybox}
\def\BibTeX{{\rm B\kern-.05em{\sc i\kern-.025em b}\kern-.08em
    T\kern-.1667em\lower.7ex\hbox{E}\kern-.125emX}}
\begin{document}

\title{How to Quantify the Security Level of Embedded Systems? A Taxonomy of Security Metrics\\}

\author{\IEEEauthorblockN{1\textsuperscript{st} Ángel Longueira-Romero, 2\textsuperscript{nd} Rosa Iglesias, 3\textsuperscript{rd} David Gonzalez}
\IEEEauthorblockA{\textit{Industrial Cybersecurity} \\
\textit{Ikerlan Technology Research Centre (BRTA)}\\
Arrasate/Mondragón, Spain \\
\{alongueira, riglesias, dgonzalez\}@ikerlan.es}
\and
% \IEEEauthorblockN{2\textsuperscript{nd} Rosa Iglesias}
% \IEEEauthorblockA{\textit{Industrial Cybersecurity} \\
% \textit{Ikerlan Technology Research Centre (BRTA)}\\
% Arrasate/Mondragón, Spain \\
% 0000-0003-1036-3035}
\and
\IEEEauthorblockN{4\textsuperscript{th} Iñaki Garitano}
\IEEEauthorblockA{\textit{Dept. of Electronics and Computing} \\
\textit{Mondragon Unibertsitatea}\\
Arrasate/Mondragón, Spain \\
igaritano@mondragon.edu}
}

\maketitle
\begin{abstract}
Embedded Systems (ES) development has been historically focused on functionality rather than security, and today it still applies in many sectors and applications. However, there is an  increasing number of security threats over ES, and a successful attack could have economical, physical or even human consequences, since many of them are used to control critical applications. A standardized and general accepted security testing framework is needed to provide guidance, common reporting forms, and the possibility to compare the results along the time. This can be achieved by introducing security metrics into the evaluation or assessment process. If carefully designed and chosen, metrics could provide a quantitative, repeatable and reproducible value that would reflect the level of security protection of the ES. This paper analyzes the features that a good security metric should exhibit, introduces a taxonomy for classifying them, and finally, it carries out a literature survey on security metrics for the security evaluation of ES. In this review, more than 500 metrics were collected and analyzed. Then, they were reduced to 169 metrics that have the potential to be applied to ES security evaluation. As expected, the $77.5 \%$ of them is related exclusively to software, and only the $0.6 \%$ of them addresses exclusively hardware security. This work aims to lay the foundations for constructing a security evaluation methodology that uses metrics to quantify the security level of an ES.
\end{abstract}


% =================================================================================
% === I. INTRODUCTION =============================================================
% =================================================================================
\section{Introduction}

The current society is driven by Embedded Systems (ES). All kinds of industries, including automotive, healthcare, avionics, telecommunications, and military defense, take advantage of ES to sense and measure the environment, control different processes and perform all types of operations \cite{embeddedSystemsDesign_Marwedel, embeddedSystemsSecurity_Kleidermacher}. The development of Industry 4.0 heavily depends on the characteristics of these ES since they represent the intelligence needed for automation and digitalization.

As stated, there is clearly an increasing number of security threats over ES \cite{embeddedSystemsSecurity_Kleidermacher} and countermeasures are being applied to protect them against known threats. As such, there are some questions that arise and cannot be answered yet: Do the countermeasures work properly? Is their coverage enough to undermine threats? And, are they correctly implemented? In this situation, a standardized and general accepted security testing methodology is needed to provide guidance, confidence, structured procedures, common reporting forms, and the possibility to compare results along time and among different systems. This last factor can be achieved by introducing security metrics into the evaluation or assessment process. Furthermore, if carefully designed and chosen, metrics would provide a quantitative, repeatable and reproducible value that would reflect the level of security protection of the ES. These metrics can shed some light on measuring the degree of confidence in the effectiveness of the implementation of security measures in an ES. Metrics would allow the comparison of different evaluations of the same device (same hardware, same software, and same versions of both) over time, and even evaluations of different versions or devices.

Security has been widely analyzed in Information Technology (IT) systems, but ES have some particular properties, such as resource limitation, physical protection difficulty, use of insecure industrial protocols, cost limitations, among others, which make it very challenging to secure them. Countermeasures, such as encryption, authentication, firewalls, secure protocols (e.g., TLS/SSL), intrusion detection, and intrusion prevention systems are commonly used in networks and PCs, but these cannot be directly applied to solve security issues of ES. On the other hand, most ES are not upgraded regularly, but have much longer lifespan and a successful attack can be easily escalated on other devices of the same kind. The aim of this paper is to provide a broader set of criteria (metrics) as a part of a more comprehensive process for evaluating the security of ES, that is, that allows to measure the degree of confidence in the effectiveness of protections and countermeasures.

This paper structure is as follows: Section II describes current security metrics in literature and standards, raised issues, and proposed taxonomies. Section III describes the characteristics of a good security metric. In Section IV, a new taxonomy for metrics based on the characteristics of ES is proposed, along with an analysis of the metrics collected in the literature. In Section V, conclusions and future work are introduced.

% =================================================================================
% === II. Previous Work on Security Metrics =======================================
% =================================================================================
\section{Previous Work on Security Metrics}
    In general, most evaluation criteria reflect two different aspects of security: functionality and assurance. Functionality refers to the means that are used, e.g., types of access control, or authentication mechanisms. On the other hand, security assurance is mostly based on development practices, documentation, analysis and configuration, and less in testing because it is more difficult, time-consuming and costly. Thus, current approaches are more focused on subjective than objective measures, that is, they are mainly based on the reviews of a group of evaluators. Without good metrics and corresponding evaluation methods, security cannot be accurately diagnosed and quantitatively measured.
    
     First, to understand what a security metric is, it is important to distinguish between ``metric" and ``measurement". A measurement is a concrete and objective attribute that provides a single-point-in-time view of a specific and discrete factor; whereas, a metric is generated from the analysis of the raw data provided by measurements \cite{Zeb_QuantitativeSecurityMetrics_2018}. An example of measurement would be the number of spams detected. A corresponding metric would be the number of spams detected last month compared to the number of spam detected during this month.
    
    This section highlights the importance of security metrics, describes the existing metrics proposed in international standards and analyzes the proposed taxonomies.

    \subsection{Metrics in the literature}
        % Surveys
       In \cite{whySecurityMetrics_Atzeni}, the authors conclude that measurements and metrics are necessary to justify investments in security and to manage security. On the other hand, in \cite{securityMetricsSurvey_Pendleton}, Pendleton \textit{et al.} provide a comprehensive survey on quantitative system security metrics. In their work, they analyze software, network, economic, and effectiveness security metrics. To classify the surveyed metrics, a hierarchical ontology is proposed, and then, each metric is briefly described. A survey that focuses on model-based quantitative security metrics is presented in \cite{modelbasedNetworkMetrics_Ramos}. The authors review the state of the art of network security metrics in depth, more specifically, in the realm of model-based quantitative metrics. They present a complete and thorough review of metrics proposals, remarking the pros and cons of each metric. 
        
        In 2019, it is highlighted the lack of consistent security metrics towards a common certification approach \cite{MatheuSara}. These metrics, together with a suitable certification approach would help to assess and compare the security of different ES, and would increase the trust of the end user. 

    \subsection{Metrics in standards and methodologies}
        Over time, standards have also proposed the use of security metrics to evaluate the degree of compliance achieved.
        
        % IEC 62443
    	The IEC 62443-1-3 standard \cite{ISA_IEC62443_2009} includes cybersecurity conformance metrics for managing a security program throughout the life cycle of an industrial automation control system. These metrics provide evidence of conformance for verifiability, completeness, and accuracy. They can also be of assistance to others for assessing the cyber robustness of their control system solutions. This standard also provides a list with the characteristics of a good metric, a metrics development and implementation process, and the steps to be followed for creating metrics. The 62443-4-1 standard also includes some examples of metrics that can be used at the component level, that is, for ES, network components, host software and applications, e.g., functional security, deployment security and current backlog. In Annex A of \cite{ISA_IEC62443_2009}, possible metrics are described to be used in the case of highest maturity level for an organization. Some of these metrics are: results of static code analysis, analysis of attack surface and deviations from Secure Coding Guidelines. They show the effectiveness of the development process with measurable improvements.
    
    	% COMMON CRITERIA
    	The Common Criteria for Information Technology Security Evaluation (Common Criteria or CC for short) is an international standard (ISO/IEC 15408) for computer security certification \cite{commonCriteriaGeneralModel}. CC is a framework which provides assurance that the process of specification, implementation and evaluation of a computer security product has been conducted in a rigorous and standard and repeatable manner at a level that is commensurate with the target environment for use. To describe the rigor and depth of an evaluation, the CC defines the Evaluation Assurance Levels (EALs) as an increasing scale. CC list seven levels, from EAL1 (the most basic one) to EAL7 (the most stringent security level). It is important to notice that the EAL levels do not measure security itself. Instead, the emphasis is given to “functional testing”, confirming the overall security architecture and design, and performing some testing depending on the EAL to be achieved. The testing methodology defined in CC for each EAL is underspecified. As a result, evaluations can vary from one laboratory to another, and results are difficult to be reproduced. It can be said that the highest level means that more testing was done, but not necessarily that the product is more secure.
    	
    	The  Common Criteria for Information Technology Evaluation (CEM) \cite{commonCriteriaCEM} suggests the use of an effective and measurable life-cycle model that addresses the development and maintenance processes and that uses some quantitative evaluation. It explicitly mentioned three metrics: source code complexity, defect density (errors per size of code), and mean time to failure. The CEM also suggests that generated documentation should include numerical values of each metric used, as well as the actions taken as a result of the measurements, but there is no reference about how to compute these metrics.
    
    	% NIST
    	The National Institute of Standards and Technology (NIST) has done a lot of research on security metrics, and has proposed nine security metrics for three different aspects: (1) implementation, (2) effectiveness/efficiency, and (3) impact \cite{metrics_NIST1}. NIST presents its security metrics taxonomy in \cite{metrics_NIST2} and \cite{metrics_NIST3}. The taxonomy is comprehensive, presenting three security categories, (1) management, (2) technical, and (3) operational; however these metrics address security at the organization level, and do not apply to ES. A similar approach is proposed in the security framework developed by the Industrial Internet Consortium (IIC) \cite{iic_securityFramework}. According to IIC, metrics should be used from the moment systems are conceived, from their design, creation to operation. Such metrics help identify security problems early and assist in faster and more efficient management and governance.
    	
        % OSSTMM
    	OSSTMM (Open Source Security Testing Methodology Manual) \cite{isecomOSSTMM_PDF} describes a complete security auditing methodology, offering fairly good tools to report the result set. It is designed to audit the operational security of physical locations, workflows, human security testing, physical security testing, wireless security testing, telecommunication security testing, data networks security testing and compliance. OSSTMM defines the \textit{rav}, a scale measurement of the attack surface. In this scale, 100 \textit{rav} means perfect balance and a lower value indicates insufficient controls and therefore a greater attack surface. A value greater than 100 \textit{rav} shows more controls than necessary, which might be inefficient and sometimes a problem, since excessive controls often increase complexity and maintenance issues. The \textit{rav} does not measure risk for an attack surface, rather it enables the measurement of it.    	

    \subsection{Current Issues with Security Metrics}
        There are many recurrent problems associated with security metrics that are identified in the literature. In \cite{criticalSecurityIndicators_Rudolph} and \cite{Sentilles2018WhatDW}, the authors identified that hardly any security indicator has a solid theoretical foundation or empirical evidence in support of the claimed correlation. They also highlighted that many security metrics lack an adequate description of the scale, unit and reference values for comparing and interpreting results, to name a few. Moreover, only a few implementations or programs were available, and only one performed some kind of benchmarking or comparison to similar metrics. The authors concluded to a large extent that the information provided in the papers is insufficient to directly apply the method, especially by non-experts.

        Even though metrics and methodologies have been proposed in the literature, it is difficult to understand whether they are applicable in a given context, and how to use them. It seems reasonable that future research should be focused on the development of a convincing theoretical foundation, empirical evaluation, and systematic improvement of existing approaches.
        As stated in this section, there is clearly a growing interest in adopting and using security metrics, but this interest confronts with a lack of widely accepted solutions.

    \subsection{Taxonomies for security metrics}
        In \cite{Savola_2009, taxonomy_Savola}, Savola reviews the most common taxonomies for security metrics, and also provides a new model to taxonomize security metrics for technical systems to systematize and organize development activities. Although there are many taxonomies for security metrics proposed in the literature, the most common classification for metrics divides them as 
        (1) organizational (to describe and track how effectively organizational programs and processes achieve cybersecurity goals); 
        (2) technical (to describe and compare technical objects -e.g., algorithms, specifications and architectures- and to indicate the security level a specific system exhibits); and 
        (3) operational (to describe and manage the risk to an operational environment, including used systems and operating practices) \cite{quantitativeSecurityMetrics_Sanders_2014, securityMeasuring_Stolfo}.

% =================================================================================
% === III. Features of a Good Metric ==============================================
% =================================================================================
\section{Features of a Good Metric}
    Metrics are important for quantifying the security level achieved; however, computing metrics can be exhausting and time-consuming resulting in not being practical. Furthermore, it is not resource-effective to measure everything, thus using the appropriate metrics is critical. In this section, a conceptual mapping of related criteria, based on the most three relevant research works is described. This map will help in identifying security metrics that show the degree of confidence in security effectiveness.

    Measuring security, both qualitatively and quantitatively, is not an easy task \cite{whySecurityMetrics_Atzeni, whySecurityTesting_Herbert, weakHypothesis_Verendel}. It is a long-standing open problem to the research community and it is of practical importance to software industry today. Suitable metrics are needed to achieve a foundational science to guide system design and development, and to reveal the safety, security, and possible fragility of complex systems. Using metrics allows systems to be compared and evaluated, so we can increase accountability, demonstrate compliance and determine how much our investments in products and processes are making our systems more secure \cite{securityMeasuring_Stolfo, securityHard}.

    Some authors have tried to characterize which criteria should meet a good metric. Among their efforts, three main ideas can be found in the literature: (1) the SMART criterion \cite{smart_Doran}, (2) the PRAGMATIC criterion \cite{pragmaticMetrics_Krag} and (3) the characteristics identified in the work carried out by Savola \cite{qualityOfMetrics_Savola}.

    Since Doran proposed the ``SMART" (Specific, Measurable, Attainable, Repeatable, Time-dependent) criteria in \cite{smart_Doran}, it has been widely used, developed and extended, and also adapted for security metrics development \cite{securityMetrics_Jelen}. On the other hand, Krag \textit{et al.} \cite{pragmaticMetrics_Krag} propose a set of nine criteria for assessing and selecting metrics, using the acronym ``PRAGMATIC" (Predictive, Relevant, Actionable, Genuine, Meaningful, Accurate, Timely, Independent, Cheap).

    In \cite{qualityOfMetrics_Savola, feasibilityMetrics_Savola}, Savola analyzes the literature to identify the quality criteria and dimensions of security metrics and measurement process. In \cite{feasibilityMetrics_Savola}, the pre-existing goodness criteria are listed. Afterwards, in \cite{qualityOfMetrics_Savola}, Savola takes research one step further by conducting expert surveys and interviews to 141 security experts from 21 different countries, and extracting 19 quality criteria. According to his results, correctness, measurability, and meaningfulness are the main quality criteria along with usability.

    The quality criteria that Savola proposes are intended to support human decision-making in every step of the life cycle, e.g., security engineering activities at design and development practices or in security management activities at run time. In his work, Savola concludes that there is no security approach that would allow the measurement of security as an universal property, since there is no security metric that could fulfill all the quality criteria, so security metrics cannot be used to measure security as a whole. He finally points out that there is a need for widely-accepted approaches for security measuring.

\begin{figure*}
      \begin{center}
      \includegraphics[width=6.8in]{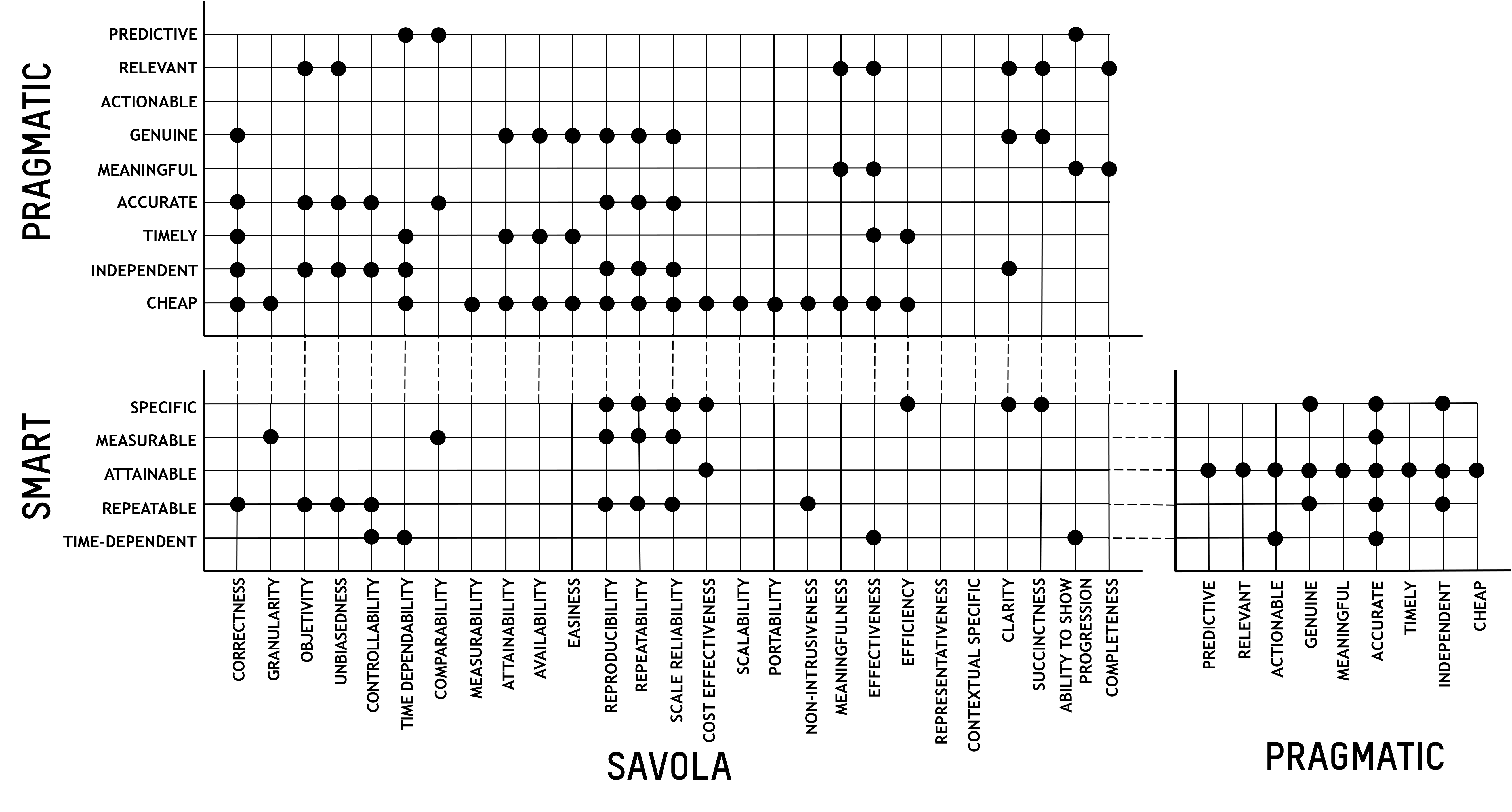}
      \caption{Mapping of criteria for identifying a good metric, according to the SMART, and PRAGMATIC criteria; and the survey carried out by Savola. Each black circle represents a common point between each one of the properties.}
      \label{fig:goodMetricFeaturesComparison}
      \end{center}
    \end{figure*}

    Figure \ref{fig:goodMetricFeaturesComparison} shows the mapping among the three different concepts explained in each criterion. This mapping was carried out using the definition and the data given by the authors, to find the common points among them. This figure can be better visualized at \url{https://embeddedsecuritymetrics.github.io/}.

    % =======
    % FIG. 01
    % =======

    Savola's approach classified properties into 19 items; whereas the other two criteria have 9 and 5 criteria resulting in a higher mapping ratio. With this figure it is not possible to deduce which property is more important in any context, but it is possible to conclude that cost, accuracy, genuineness and repeatability are the most common properties in all three criteria. 

    It should be noted that none of these criteria explicitly mentions that metrics should be quantitative. This may be because, depending on the context, a nominal scale could be used instead. Some authors explicitly avoid this approach and prefer metrics to be quantitative \cite{metrics_Jaquith}. In this mapping, many criteria have associated the cheap feature. This is because there are properties that directly affect the cost of computing a metric (the more properties it meets, the more expensive it is).

    On the other hand, there are two properties identified by Savola - \textit{i.e.}, contextual specific and representativeness - that do not have any correspondence with the rest of the criteria.

    In the context of ES, the quality criteria that Security Metrics (SM) should address are (in no particular order):
    \begin{itemize}
        \item \textbf{Comparability:} SM should support comparison of the targets that they represent.
        \item \textbf{Cost effectiveness:} measurement gathering and approaches should be cost effective.
        \item \textbf{Measurability:} SM are capable of having dimensions, quantity or capacity ascertained in the ES.
        \item \textbf{Repeatability:} The same results are achieved if a measurement is repeated in the same context, with exactly the same conditions.
        \item \textbf{Reproducibility:} The same results are achieved if a measurement is repeated in the same context, with exactly the same conditions, and different persons.
    \end{itemize}

% =================================================================================
% === IV. Security Metrics and Taxonomy ===========================================
% =================================================================================

\section{Security Metrics and Taxonomy}
    In this section, the results of a literature review about security metrics, and a new taxonomy for security metrics are described.
    
    \subsection{Selecting metrics for embedded systems}
    	There is neither a single system metric nor ``one-perfect" set of metrics that could be suitable for every single case. The set of metrics that will be most suitable depends on multiple factors. For that reason, it is convenient to analyze whether the existing metrics can be applied to ES. In this research work, a systematic review of the state of the art is carried out in order to search for the currently available security metrics. The objective is to analyze to what extent the proposed metrics can be applicable to the security evaluation of ES. The details of this search are explained in the following points.
    	
    	\subsubsection{Search and selection strategy}
    	    The data sources used are online databases, conference proceedings and academic journals of IEEE Xplore, Elsevier, AMC Digital Library, Springer, along with Google Scholar search engine. The search terms included the keywords: ``security'', ``metric'', ``measure'', ``evaluation'' and ``assessment'', considering another synonyms.
    	    
    	    The inclusion criteria included:
    	    \begin{itemize}
    	        \item Security measurements or metrics are the main subjects.
    	        \item Surveys collecting security metrics were preferred.
    	        \item The paper is primarily related to measuring security.
    	    \end{itemize}
    	    The initial number of metrics analyzed was 531, obtained mainly from \cite{securityMetricsSurvey_Pendleton, modelBasedNetwork, compositeMetricsNetwork, mappingMetrics}. For each one, the following data was collected:
        	\begin{itemize}
        	    \item \textbf{Definition:} Definition of the metric given by the author.
        	    \item \textbf{Scale:} Nominal, ordinal, interval, ratio, absolute and distribution \cite{securityMetricsSurvey_Pendleton}.
        	    \item \textbf{Scope:} User, software, hardware, device or organization.
        	    \item \textbf{Automation:} Automatic, semi-automatic or manual.
        	    \item \textbf{Measurement:} Static or dynamic measurement.
        	\end{itemize}

        \subsubsection{Filtering and exclusion criteria}
            In order to identify the subset of metrics that can be also applied to ES from that initial set, the collected metrics were filtered according to the following criteria:
            \begin{itemize}
                \item Organizational specific metrics.
                \item Network specific metrics.
                \item Metrics without a clear definition.
                \item Repeated metrics.
            \end{itemize}
            When the concept of a metric was found to be applicable with some modification, it was considered to be eligible.

        \subsubsection{Data extraction and interpretation}
            After filtering the data, the final count of metrics was 169. An interesting, albeit not very surprising fact, is that most of the metrics ($77.5 \%$ of them) were related exclusively to software - e.g., lines of code, number of functions and so on. On the other hand, only $0.6 \%$ of them were related exclusively to hardware - e.g., side-channel vulnerability factor metric; and finally, $14.8 \%$ of them could be applied to both software and hardware - e.g., the historically exploited vulnerability metric that measures the number of vulnerabilities exploited in the past. The remaining $7.1 \%$ is focused on other aspects, such as user usability. This shows that there is a clear lack of hardware security metrics in the literature, and main efforts are centered in developing software-related metrics. It is worth commenting these metrics are not only valid for ES, but could also be applied to other systems.
            
            All the data extracted from the articles, including the 169 security metrics, as well as the filtering and selection processes, are available at \url{https://embeddedsecuritymetrics.github.io/}.
            
    \subsection{A new taxonomy for security metrics}
\begin{figure}
          \begin{center}
          \includegraphics[width=3.4in]{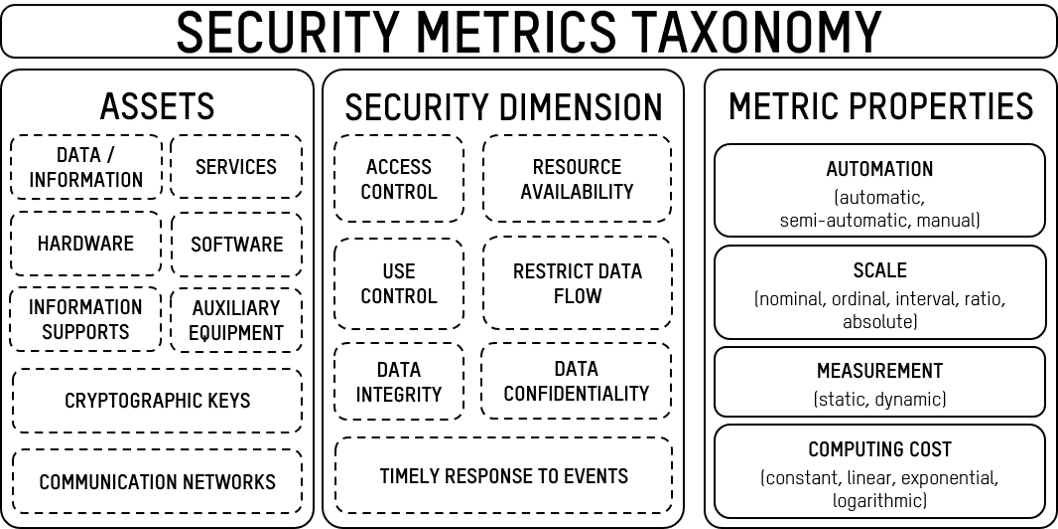}
          \caption{Taxonomy structure for security evaluation metrics for ES. Dashed line indicates AND/OR condition, and solid line AND condition.}
          \label{fig:taxonomy}
          \end{center}
        \end{figure}

        Although Section II discussed the most commonly used taxonomy for classifying security metrics, to the best of the authors knowledge, there is no taxonomy that focuses on the specific assessment needs and limitations of ES. On the other hand, the taxonomies proposed in the literature are usually at a very high level, and therefore not applicable to ES. Thus, a new taxonomy is proposed that is based on assets to be protected in ES. Figure \ref{fig:taxonomy} shows the structure of the proposed taxonomy.
        
        According to the research work in \cite{embeddedPyramid}, embedded security cannot be solved at a single level of abstraction; instead must be addressed at all abstraction levels. The main structure of the proposed taxonomy is built on three main blocks that aim to address all those abstraction levels:
        (1) Assets: elements of the ES that can be evaluated, as hardware, software, data, services, crypto keys, or communications; 
        (2) Security Dimension: which attribute is to be protected (foundational requirements from the IEC 62443); and 
        (3) Metric Properties: which are the characteristic parameters of the metric (such as automation, or computing cost).
        It is important to notice that a metric might cover more than one asset, and/or more than one security dimension, but at least, one value for each one have to be assigned (dashed line in Figure \ref{fig:taxonomy}). Nevertheless, all values in metric properties are needed for each metric, as they represent the way the metric behaves, the cost associated to a metric and the information it returns (solid line in Figure \ref{fig:taxonomy}).

        % =======
        % FIG. 02
        % =======

        The use of this taxonomy can be further explained using the Side-Channel Leakage (SCL) metric as an example. This metric uses statistical tests to measure if a device is prone to a side channel attack by comparing two sets of data; a random one and a fixed one. The SCL metric can be classified as follows:
            \begin{itemize}
                \item Assets: system hardware.
                \item Security dimension: data confidentiality.
                \item Metric properties:
                \begin{itemize}
                    \item Automation: semi-automation.
                    \item Scale: nominal (yes/no).
                    \item Measurement: dynamic.
                    \item Computing cost: constant.
                \end{itemize}                
            \end{itemize}

        As can be seen, the main feature of the proposed taxonomy is that considers all assets in ES. Unlike other proposals in the literature that are at a high level - e.g., enterprise level - this one is specifically designed to evaluate all aspects that comprise ES. Furthermore, it includes not only the CIA triad (i.e. Confidentiality, Integrity and Availability), but also other dimensions, such as, access control, use control, restrict data flow and timely-response to events. These dimensions are foundational requirements in IEC 62443 that need to be evaluated in ES.

% =================================================================================
% === V. Conclusion ===============================================================
% =================================================================================
\section{Conclusions and Future Work}
Security metrics are not standardized, and they would help in quantifying the security level of ES, that is, in giving the degree of confidence in the effectiveness of protections and countermeasures. This research work proposes a new taxonomy for security metrics, which takes into account the different assets of ES. Since it is critical to identify proper and non-exhausting security metrics for this security assurance, a conceptual mapping of criteria for choosing the most appropriate metrics is given. On the other hand, the literature has been analyzed, and 169 security metrics have been identified. $77.5 \%$ of them are dedicated exclusively to evaluating software security, while only $0.6 \%$ of them are oriented to hardware evaluation. This difference clearly shows the wider adoption by the community of software-related security metrics.

As a future work, a reduced set of metrics from those that have been analyzed will be selected, based on the 5 proposed criteria, refined, and applied to a real ES. This will be the foundation for proposing a security assessment methodology specifically designed for ES, which will provide qualitative and quantitative results. These results will give trust to end-users, make it possible to monitor the evolution of the security level of a device over time, and to compare different versions or devices.

\bibliographystyle{ieeetr}
\bibliography{Bibliography}

\begin{thebibliography}{10}

\bibitem{embeddedSystemsDesign_Marwedel}
P.~Marwedel, {\em Embedded System Design: Embedded Systems Foundations of
  Cyber-Physical Systems}.
\newblock Springer, 01 2010.

\bibitem{embeddedSystemsSecurity_Kleidermacher}
D.~Kleidermacher and M.~Kleidermacher, {\em Embedded Systems Security:
  Practical Methods for Safe and Secure Software and Systems Development}.
\newblock Newton, MA, USA: Newnes, 1st~ed., 2012.

\bibitem{Zeb_QuantitativeSecurityMetrics_2018}
T.~{Zeb}, M.~{Yousaf}, H.~{Afzal}, and M.~R. {Mufti}, ``A quantitative security
  metric model for security controls: Secure virtual machine migration protocol
  as target of assessment,'' {\em China Communications}, vol.~15, no.~8,
  pp.~126--140, 2018.

\bibitem{whySecurityMetrics_Atzeni}
A.~Atzeni and A.~Lioy, ``Why to adopt a security metric? a brief survey,'' {\em
  Advances in Information Security}, vol.~23, pp.~1 -- 12, 2006.

\bibitem{securityMetricsSurvey_Pendleton}
M.~Pendleton, R.~Garcia-Lebron, J.-H. Cho, and S.~Xu, ``A survey on systems
  security metrics,'' {\em ACM Comput. Surv.}, December 2016.

\bibitem{modelbasedNetworkMetrics_Ramos}
A.~{Ramos}, M.~{Lazar}, R.~H. {Filho}, and J.~J. P.~C. {Rodrigues},
  ``Model-based quantitative network security metrics: A survey,'' {\em IEEE
  Communications Surveys Tutorials}, vol.~19, pp.~2704--2734, Fourthquarter
  2017.

\bibitem{MatheuSara}
S.~N. {Matheu}, J.~L. {Hernandez-Ramos}, and A.~F. {Skarmeta}, ``Toward a
  cybersecurity certification framework for the internet of things,'' {\em IEEE
  Security Privacy}, vol.~17, no.~3, pp.~66--76, 2019.

\bibitem{ISA_IEC62443_2009}
I.~E. Commission, ``{Security for industrial automation and control systems –
  Network and system security: System security compliance metrics},'' standard,
  IEC.

\bibitem{commonCriteriaGeneralModel}
{CC}, ``{The Common Criteria for Information Technology Security Evaluation -
  Introduction and General Model}.''

\bibitem{commonCriteriaCEM}
{CC}, ``{Common Methodology for Information Security Technology Evaluation}.''

\bibitem{metrics_NIST1}
E.~Chew, M.~Swanson, K.~Stine, N.~Bartol, A.~Brown, and W.~Robinson,
  ``Performance measurement guide for information security (draft),'' July
  2008.

\bibitem{metrics_NIST2}
M.~Swanson, ``Security self-assessment guide for information technology
  systems,'' November 2001.

\bibitem{metrics_NIST3}
M.~Swanson, N.~Bartol, J.~Sabato, J.~Hash, and L.~Graffo, ``Security metrics
  guide for information technology systems,'' July 2003.

\bibitem{iic_securityFramework}
{IIC}, ``{Industrial Internet of Things Volume G4: Security Framework},'' 2016.

\bibitem{isecomOSSTMM_PDF}
{ISECOM}, {\em {T}he {O}pen {S}ource {S}ecurity {T}esting {M}ethodology
  {M}anual (OSSTMM)}.
\newblock ISECOM, 2010.

\bibitem{criticalSecurityIndicators_Rudolph}
M.~{Rudolph} and R.~{Schwarz}, ``A critical survey of security indicator
  approaches,'' in {\em 2012 Seventh International Conference on Availability,
  Reliability and Security}, pp.~291--300, Aug 2012.

\bibitem{Sentilles2018WhatDW}
S.~Sentilles, E.~Papatheocharous, and F.~Ciccozzi, ``What do we know about
  software security evaluation? a preliminary study,'' in {\em QuASoQ@APSEC},
  2018.

\bibitem{Savola_2009}
R.~{Savola}, ``A security metrics taxonomization model for software-intensive
  systems,'' {\em Journal of Information Processing Systems}, 2009.

\bibitem{taxonomy_Savola}
R.~{Savola}, ``Towards a security metrics taxonomy for the information and
  communication technology industry,'' in {\em International Conference on
  Software Engineering Advances (ICSEA 2007)}, pp.~60--60, Aug 2007.

\bibitem{quantitativeSecurityMetrics_Sanders_2014}
W.~H. {Sanders}, ``Quantitative security metrics: Unattainable holy grail or a
  vital breakthrough within our reach?,'' {\em IEEE Security Privacy}, Mar
  2014.

\bibitem{securityMeasuring_Stolfo}
S.~{Stolfo}, S.~M. {Bellovin}, and D.~{Evans}, ``Measuring security,'' {\em
  IEEE Security Privacy}, May 2011.

\bibitem{whySecurityTesting_Herbert}
S.~M. Bellovin, ``On the brittleness of software and the infeasibility of
  security metrics,'' {\em IEEE Security \& Privacy}, vol.~4, no.~4,
  pp.~96--96, 2006.

\bibitem{weakHypothesis_Verendel}
V.~Verendel, ``Quantified security is a weak hypothesis: A critical survey of
  results and assumptions,'' in {\em Proceedings of the 2009 Workshop on New
  Security Paradigms Workshop}, NSPW '09, (New York, NY, USA), pp.~37--50, ACM,
  2009.

\bibitem{securityHard}
S.~{Pfleeger} and R.~{Cunningham}, ``Why measuring security is hard,'' {\em
  IEEE Security Privacy}, July 2010.

\bibitem{smart_Doran}
G.~T. Doran, ``There's a s.m.a.r.t. way to write managements's goals and
  objectives.,'' {\em Management Review}, vol.~70, no.~11, pp.~35--36, 1981.

\bibitem{pragmaticMetrics_Krag}
{W. Krag Brotby and Gary Hinson}, {\em {PRAGMATIC Security Metrics. Applying
  Metametrics to Information Security}}.
\newblock Auerbach Publications, 01 2013.

\bibitem{qualityOfMetrics_Savola}
R.~M. Savola, ``Quality of security metrics and measurements,'' {\em Computers
  \& Security}, vol.~37, pp.~78 -- 90, 2013.

\bibitem{securityMetrics_Jelen}
G.~Jelen, ``{SSE-CMM Security Metrics},'' in {\em National Institute of
  Standards and Techonology (NIST) and Computer System Security and Privacy
  Advisory Board (CSSPAB) Workshop}, June 2000.

\bibitem{feasibilityMetrics_Savola}
R.~Savola, ``On the feasibility of utilizing security metrics in
  software-intensive systems,'' {\em International Journal of Computer Science
  and Network Security}, vol.~10, pp.~230--239, 01 2010.

\bibitem{metrics_Jaquith}
A.~Jaquith, {\em Security Metrics: Replacing Fear, Uncertainty, and Doubt}.
\newblock Addison-Wesley Professional, 01 2007.

\bibitem{modelBasedNetwork}
A.~{Ramos}, M.~{Lazar}, R.~H. {Filho}, and J.~J. P.~C. {Rodrigues},
  ``Model-based quantitative network security metrics: A survey,'' {\em IEEE
  Communications Surveys Tutorials}, vol.~19, no.~4, pp.~2704--2734, 2017.

\bibitem{compositeMetricsNetwork}
S.~Yusuf~Enoch, J.~Hong, M.~Ge, and D.~Kim, ``Composite metrics for network
  security analysis,'' vol.~2017, pp.~137--160, 02 2017.

\bibitem{mappingMetrics}
P.~Morrison, D.~Moye, and L.~L. Williams, ``Mapping the field of software
  security metrics,'' 2014.

\bibitem{embeddedPyramid}
D.~D. {Hwang}, P.~{Schaumont}, K.~{Tiri}, and I.~{Verbauwhede}, ``Securing
  embedded systems,'' {\em IEEE Security Privacy}, vol.~4, no.~2, pp.~40--49,
  2006.

\end{thebibliography}

\end{document}